\documentclass[conference]{IEEEtran}
\IEEEoverridecommandlockouts
\usepackage{cite}
\usepackage{amsmath,amssymb,amsfonts}
\usepackage{amsthm}
\usepackage{algorithm}
\usepackage{algorithmic}
\usepackage{graphicx}
\usepackage{textcomp}
\usepackage{xcolor}
\usepackage{csquotes}
\usepackage{subcaption}
\usepackage{dsfont}
\usepackage{cite}
\usepackage{balance}
\usepackage{adjustbox}
\usepackage{bbm}
\usepackage{booktabs}

\usepackage{varwidth}
\usepackage{soul}
\usepackage{array, multirow}
\usepackage{hyperref}

\begin{document}
\title{Optimized Control-Centric Communication in Cooperative Adaptive Cruise Control Systems}
\author{Mahdi Razzaghpour, Shahriar Shahram, Rodolfo Valiente, Mahdi Zaman, Yaser P. Fallah
}
\maketitle

\begin{abstract}
In this study, we explore an innovative approach to enhance cooperative driving in vehicle platooning systems through the use of vehicle-to-everything (V2X) communication technologies. As Connected and Autonomous Vehicles (CAVs) integrate into increasingly dense traffic networks, the challenge of efficiently managing communication resources becomes crucial. Our focus is on optimizing communication strategies to support the growing network of interconnected vehicles without compromising traffic safety and efficiency. We introduce a novel control-aware communication framework designed to reduce communication overhead while maintaining essential performance standards in vehicle platoons. This method pivots from traditional periodic communication to more adaptable aperiodic or event-triggered schemes. Additionally, we integrate Model-Based Communication (MBC) to enhance vehicle perception under suboptimal communication conditions. By merging control-aware communication with MBC, our approach effectively controls vehicle platoons, striking a balance between communication resource conservation and control performance. The results show a marked decrease in communication frequency by 47\%, with minimal impact on control accuracy, such as less than 1\% variation in speed. Extensive simulations validate the effectiveness of our combined approach in managing communication and control in vehicle platoons, offering a promising solution for future cooperative driving systems.
\end{abstract}

\section{Introduction} \label{sec::intro}
\noindent Cooperative Adaptive Cruise Control (CACC), utilizing Vehicle-to-Vehicle (V2V) communication, is key to improving traffic dynamics by promoting string stability and enabling reduced spacing between vehicles \cite{double_throughput, Safety_ECC}. At the heart of distributed Multi-Agent Systems (MASs) is the critical role of information sharing, which is essential for achieving a comprehensive understanding of the situation. However, excessive use of communication channels can lead to congestion, characterized by longer latency periods, increased instances of packet loss, and decreased throughput. These issues can significantly affect the stability, efficiency, and dependability of the system \cite{Information_Dissemination}. Therefore, when designing distributed control systems for MASs, it is imperative to balance achieving the targeted control outcomes with the judicious use of limited communication and computational resources. This approach ensures that system performance is optimized without overwhelming the available infrastructure.

\begin{figure}
    \centering
    \includegraphics[width=\linewidth,trim={0mm 5mm 0mm 0mm},clip]{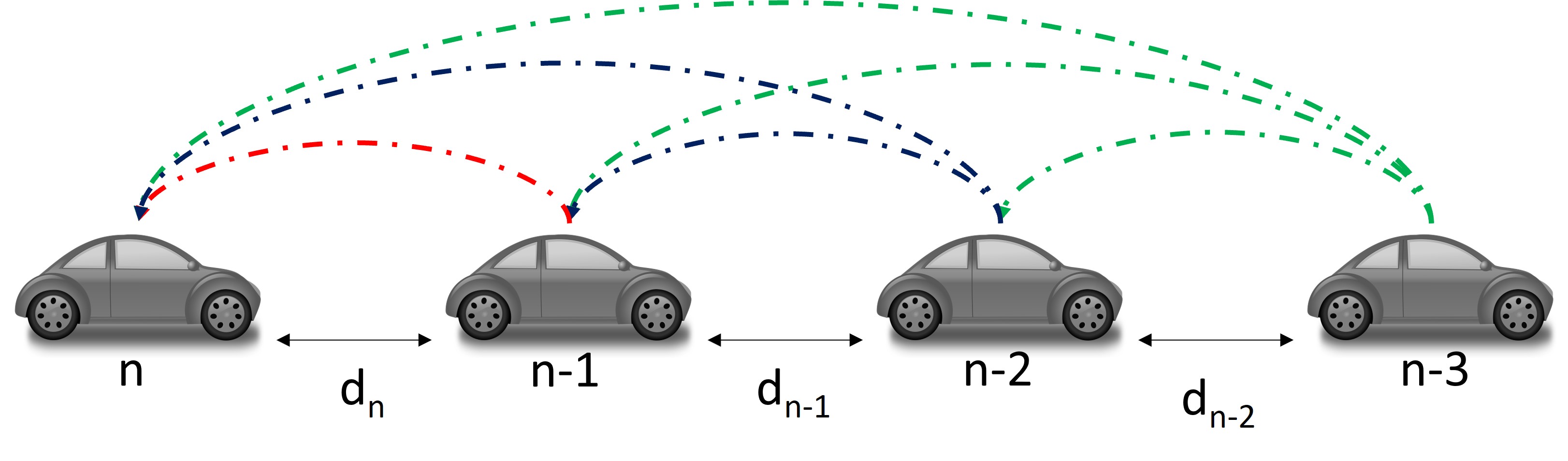}
    \vspace{-0.15in}
    \caption{An explanation of the communication structure used by vehicles is presented, where dashed lines represent the exchange of information between them. The term $d_i$ is used to describe the gap between the $n^{th}$ vehicle and the one directly in front of it.}
    \label{fig:diagram}
    \vspace{-0.15in}
\end{figure}

This abstract is centered on the concept of Model-Based Communication (MBC), a novel strategy aimed at scaling communication more effectively, with a particular emphasis on alleviating channel congestion \cite{model_based_communication}. MBC differentiates itself by employing a content structure optimized for conveying information pertinent to the integrated dynamics of both vehicle and driver behavior. This approach deviates from the traditional content structure of Basic Safety Messages (BSM) as specified in the J2735 standard \cite{saej2735}. Such a strategy becomes increasingly pertinent in the context of diverse vehicle dynamic modeling methods. Among these methods, non-parametric Bayesian inference techniques, such as Gaussian Processes (GP), are recognized for their considerable potential and effectiveness.

Unlike the commonly used Time-Triggered Communication (TTC) in current vehicle platoon control studies, which relies on consistent data exchange intervals without adapting to measurement changes, MBC introduces a more flexible and responsive approach. Traditional TTC operates on preset communication intervals, often overlooking real-time vehicle conditions, resulting in suboptimal use of communication resources, a critical issue in CACC systems. In contrast, MBC advocates for a communication method that is more attuned to the control requirements. This approach schedules transmission times based on the actual output data, aiming to achieve a harmonious balance between communication efficiency and control accuracy. Focusing on the specific demands of the control system, MBC has the potential to facilitate more effective and performance-driven communication strategies in the context of vehicle platooning.

The design of an effective Event-Triggered Communication (ETC) system for vehicle platoons is essential due to the inherent balance required between the control performance of the platoon and the usage of communication resources. This paper details the development of such a system and offers a fresh viewpoint on modeling the interactions among various elements of a vehicle platoon to enhance its overall performance. Key contributions of this paper include:
\begin{itemize}
    \item Our paper presents a novel communication solution that integrates ETC with MBC for the collaborative control of vehicle platoons.
    \item We outline a comprehensive ETC framework designed for distributed application in vehicle platoons. This approach notably decreases the average rate of communication while only minimally impacting the effectiveness of control performance. 
\end{itemize}

\section{Related Work} \label{sec::Related}
\noindent This section delves into the management of the collective behavior of multiple CAVs, which hinges on the vehicles' shared understanding of their respective states, such as the distance between vehicles and their speeds. This mutual awareness is achieved through a combination of inter-vehicle sensing and communication. Specifically, we will explore one of the key applications of cooperative driving, CACC, along with V2V communication.
\subsection{\textbf{Cooperative Adaptive Cruise Control (CACC)}}
\noindent For CACC systems to be effective, they need to be robust against unique scenarios like vehicles abruptly joining the platoon or sudden braking by lead vehicles \cite{8370701}. The advanced and accurate data provided by V2V communication enables CACC vehicles to closely follow the vehicle ahead, even at significantly reduced distances. This not only improves user acceptance but can also markedly enhance lane capacity and traffic flow dynamics. Studies have shown that vehicle platooning can significantly contribute to solving various transportation challenges \cite{Impact_CACC_Shladover, Effects_CACC_stability}. A key aspect of a successful CACC system is maintaining minimal spacing error, the deviation from the desired gap between vehicles. Keeping this error small is crucial for minimizing collision risks and reaping the benefits of platooning, such as reduced fuel consumption and increased traffic throughput \cite{fuel}.

Imperfections in communication can greatly impact the efficacy of CACC systems. Issues such as prolonged communication delays or insufficient transmission rates can disrupt string stability and various other performance aspects, especially concerning the maintenance of a specific time gap. Consequently, it is crucial to have a sufficiently high number of transmissions over time and minimal communication delays to ensure the desired behavior in vehicle platooning is achieved \cite{8025403}.

\subsection{\textbf{Vehicle-to-Vehicle (V2V) Communication}}
\noindent The exchange of information is crucial for the effective deployment of vehicle platoons, facilitating the implementation of control actions based on current road and traffic situations. Numerous studies have investigated the influence of communication networks on the performance of platoons \cite{7997746,V2V_platooning}. A significant limitation of TTC is its rigidity and limited scalability. This section introduces ETC and MBC as solutions that offer flexibility and scalability. Within the framework of the Cellular Vehicle-to-Everything (C-V2X) standard, a specific lower threshold is set for the Minimum Inter-Event Time (MIET), defining the minimum duration that must elapse between two successive transmissions \cite{saej3161}. The MIET ranges from a lower limit of $100ms$ to an upper limit of $600ms$. This positive lower bound is essential to prevent Zeno behavior, which is characterized by an infinite number of events in a finite time, and to ensure the practicality of implementing an ETC system.

\vspace{0.1in}
\subsubsection{\textbf{Optimizing MAS Communication with Event-Triggered Strategies}}
Strategies based on event triggers are widely recognized for their effectiveness in optimizing the use of communication resources within MASs \cite{Lemmon2010,SEYBOTH2013245}. These approaches, which stand in contrast to the conventional TTC, focus on transmitting data only when it is required to fulfill specific control system criteria. Studies have shown that systems triggered by events demonstrate superior real-time performance compared to those operating on time triggers. For instance, research suggests a method to reduce the communication load by employing a versatile event-triggering strategy, which involves adjustable parameters tailored for each member of a vehicle platoon \cite{WEN2018341}.

In this approach, agents transmit their current status to adjacent agents only if the difference between their present state and the last communicated state exceeds a dynamically changing threshold, or when it hits the peak of the inter-event time span. To make event-triggered methods more practical, the concept of setting a minimum time gap between consecutive events has been investigated \cite{NOWZARI20191}. The event-triggered mechanism for each vehicle in the system can be characterized in the following manner:
\begin{equation}
t_{k+1} = t_k + \min \left(\tau_k, \tau\right) \text {, }
\end{equation}
where $\tau$ represents a positive constant that signifies the maximum limit of the interval between events. The value of $\tau_k$ is determined using the equation given below:
\begin{equation}
\tau_k = \inf _{t>t_k}\left\{t-t_k \mid C\left(S(t) \:,\tilde S\left(t\right)\right)>0\right\}, \quad \text { for } t \geq 0 .
\label{ETC_form}
\end{equation}
Upon receiving fresh information, every agent updates its control input and employs the newly acquired model for predictive purposes. It is important to highlight that the timing of these triggers is not coordinated across agents. In these methods, each vehicle independently operates a parallel version of the dynamics of its neighboring vehicles. In the suggested setup, vehicles rely on their self-transmitted model to decide the timing of data transmission. If the prediction from the kinematic model remains accurate since the last transmission, the vehicle will not transmit a new message.

\vspace{0.1in}
\subsubsection{\textbf{Model-Based Communication (MBC)}}
In the development of CACC systems, it's crucial to consider the inherent uncertainties related to the state and behavior of vehicles, as well as in their communication channels \cite{Gp_VNC}. Given that information from neighboring vehicles isn't always accessible, it becomes necessary for each agent to operate an estimation system. Within this framework, agents utilize a predictive model to estimate the measurements from other agents in scenarios where data packets are not received, either due to loss of packets or because an event did not prompt a transmission.

In our study, the velocity of each cooperative vehicle over time, denoted as $v_{n}(t)$, is modeled as a GP. This process is characterized by a mean function, $m_{n}(t)$, and a covariance kernel function, $\kappa_{n}(t, t^{\prime})$, as follows:
\begin{equation}
v_{n}(\mathbf{t}) \sim \mathcal{G} \mathcal{P}\left(m_{n}(\mathbf{t}), \kappa_{n}\left(\mathbf{t}, \mathbf{t}^{\prime}\right)\right).
\end{equation}
Our focus is on integrating insights derived from observed velocity data regarding the underlying function, $v_{n}(t)$, and its future projections. We assume that for each cooperative vehicle, the process mean is zero, $m_{n}(t)=0$. We use a Radial Basis Function (RBF) as the covariance kernel and consider the measurement noises to be independent and identically distributed ($i.i.d.$) following a Gaussian distribution, $\mathcal{N}(0,\,\gamma_{n, noise}^{2})$. Under these assumptions, the covariance matrix for the observed velocity of the $n^{th}$ cooperative vehicle can be expressed as follows:
\begin{equation}
\begin{aligned}
\label{kernel matrix}
 K_{n}(\boldsymbol{t},\boldsymbol{t^{\prime}})= \kappa_{n}(t,t^{\prime}) + \gamma_{n,noise}^{2}I
\end{aligned}
\end{equation}
In this context, $I$ represents the identity matrix, whose dimension matches that of the training (measured) data. The calculation of $\kappa_{n}(t,t^{\prime})$ can be performed based on the definition of the RBF, as follows:
\begin{equation}
 \kappa_{n}(t,t^{\prime})=\exp(-\frac{||t-t^{\prime}||^2}{2\gamma_{n}^{2}}).
\end{equation}
Under the previously mentioned assumptions, we can represent $\mathcal{V}_{n}^{obs}$, and the future values, $\mathcal{V}_{n}^{\ast}$, in the following way:
\begin{equation}
\left[\begin{array}{l}
\mathbf{\mathcal{V}_{n}^{obs}} \\
\mathbf{\mathcal{V}_{n}}^{*}
\end{array}\right] \sim \mathcal{N}\left(\mathbf{0},\left[\begin{array}{ll}
K_{n}(\boldsymbol{t}, \boldsymbol{t}) & K_{n}\left(\boldsymbol{t}, \boldsymbol{t^{*}}\right) \\
K_{n}\left(\boldsymbol{t^{*}}, \boldsymbol{t}\right) & K_{n}\left(\boldsymbol{t^{*}}, \boldsymbol{t^{*}}\right)
\end{array}\right]\right),
\end{equation}
In this formulation, $\boldsymbol{t}$ and $\boldsymbol{t^{*}}$ represent the time stamps associated with the sets of observation and future values, respectively. The function $K_{n}(.,.)$ is derived as per the kernel matrix described in \eqref{kernel matrix}.

\section{Preliminaries and Problem Formulation}
\noindent Reducing V2V communication significantly, without impacting the operational efficiency of vehicular platoons, poses a considerable challenge. The main task is to devise control strategies that ensure the effective performance of MASs while markedly cutting down on the overuse of communication and computational resources. Our control strategy leverages local information, specifically spacing error and velocity error, in a relative manner. This involves evaluating these parameters in relation to the state of each agent to adjust the control input of every following vehicle. The objective is to align with the lead vehicle's speed while ensuring a steady time gap between consecutive vehicles. To mitigate the impact of communication disruptions in V2V exchanges, our approach in CACC systems involves using a GP to estimate the speeds of the vehicles ahead, effectively compensating for any loss of communication.

\subsection{\textbf{Vehicle Model and Predictive Control Design}}
\noindent In our research, we examine a platoon consisting of $N_v$ vehicles. Within this platoon, $n$ represents the index of a vehicle, where $n\in\{0,1,\hdots,N_v-1\}$, and the vehicle indexed as $n=0$ is designated as the leader of the platoon, as illustrated in Figure \ref{fig:diagram}. The term $d_n$ refers to the distance separating the $n^{th}$ vehicle from the $(n-1)^{th}$ vehicle and is defined as follows:
\begin{equation}
    d_n = x_{n-1}-x_{n}-l^v_n,
\end{equation}
where $x_n$ represents the longitudinal position of the rear bumper of the $n^{th}$ vehicle, and $l^v_n$ is the length of that vehicle. The policy for the preferred spacing is established as follows:
\begin{equation} \label{d*}
    \begin{aligned}
        d^{*}_{n}(t) = \delta_n\,v_{n}(t)+d^{s}_{n}.
    \end{aligned}
\end{equation}
In the equation \eqref{d*}, $v_{n}$ denotes the velocity of the $n^{th}$ vehicle, $\delta_n$ is the time gap, and $d^s_n$ indicates the standstill distance. The difference between the actual gap and its ideal value is denoted by \(\Delta d_{n}(t) = d_{n}(t) - d^{*}_{n}(t)\), and the velocity difference between the \(n^{th}\) vehicle and the one ahead is represented by \(\Delta v_{n}(t) = v_{n-1}(t) - v_{n}(t)\). Consequently, \(\Delta \dot{d}_n\) is transformed into \(\Delta \dot{d}_n(t) = \Delta v_n(t) - \delta_n\,a_n(t)\) and \(\Delta \dot{v}_n = a_{n-1} - a_n\), where \(a_n\) is the acceleration of the \(n^{th}\) vehicle. Considering the driveline dynamics \(f_n\), the acceleration rate of vehicle \(n\) is given by \(\dot{a_{n}}(t) = -\mathnormal{f}_{n} a_{n}(t) + \mathnormal{f}_{n}u_{n}(t)\), with \(u_{n}(t)\) serving as the vehicle's control input. Defining $S_n=[\Delta d_n\,\,\,\Delta v_n\,\,\, a_n]^T$ as the state vector for the $n^{th}$ vehicle, we can represent the state-space for each vehicle as follows:
\begin{multline}  \label{css}
        \dot{S}_n(t)=A_n\,S_n(t)+B_n\,u_n(t)+D\,a_{n-1}(t)\\ \\=
        \begin{bmatrix}
            0&1&-\delta_n \\0&0&-1\\0&0&-\mathnormal{f}_{n}
        \end{bmatrix}S_n(t)+
        \begin{bmatrix}
            0\\0\\\mathnormal{f}_{n}
        \end{bmatrix}u_{n}(t)+
        \begin{bmatrix}
            0\\1\\0
        \end{bmatrix}a_{n-1}(t).
\end{multline}
For the case of $n=0$ (the leader), the term $a_{n-1}(t)$ is substituted with zero. The discrete-time state-space model, using a first-order forward time approximation, is expressed by the following equation:
\begin{multline}\label{fdss}
        S_n(k+1) =\\ (I+t_s\,A_n)\,S_n(k)+t_s\,B_n\,u_n(k)+t_s\,D\,a_{n-1}(k),
\end{multline}
where $t_s$ represents the interval of sampling time.

The system considers specific constraints on states and inputs, including limits on acceleration and input values, compliance with road speed limits, and ensuring a safe vehicle distance (noting that negative distance implies a collision, which is to be avoided). The system must consistently adhere to the following inequalities (hard constraints):
\begin{subequations} \label{bounds}
    \begin{gather}
        \label{accbound}
        a_n^{min}\leq a_n(k)\leq a_n^{max},\\ 
        \label{inputbound}
        u_n^{min}\leq u_n(k)\leq u_n^{max},\\
        \label{speedlim}
        v_n(k)\leq v^{max},\\
        d_n(k)>0.
    \end{gather}
\end{subequations}
Furthermore, to ensure passenger comfort, changes in the system input are constrained within certain limits as follows:
\begin{equation}
    \begin{gathered}
        t_s\,u_n^{min} \leq u_n(k+1)-u_n(k) \leq t_s\,u_n^{max}.
    \end{gathered}
    \label{passenger_comfort}
\end{equation}
The design problem for the Model Predictive Control (MPC) of each vehicle is formulated as follows:
\begin{multline} \label{multi_cost}
    \sum_{k=0}^{N-1}\Bigg[(\mathbf{S}_n(k)-R_n)^T\, Q_n \,(\mathbf{S}_n(k)-R_n)\\
    +\hspace{-5pt}\sum_{i=n-r}^{n-1}\hspace{-5pt}\Big[ c^d_i\,\Big(x_i(k)-x_n(k)-\hspace{-5pt}\sum_{j=i+1}^{n}\hspace{-3pt}(d^*_j(k)+l^v_j)\Big)^2\\
     \qquad \qquad \qquad  \quad  +c^v_i\,\Big(v_i(k)-v_n(k)\Big)^2\Big]\Bigg],\\
        \text{subject to: System Constrains,} \qquad \qquad
\end{multline}
In this context, $\textbf{u}_n$ refers to the inputs of the system ranging from $k=0$ to $k=N-1$. The variables $c^d_i$ and $c^v_i$ are positive coefficients, and $r$ signifies the count of preceding vehicles that are exchanging information with the $n^{th}$ vehicle.

\subsection{\textbf{The Role of Event-Triggered Conditions}}
\noindent In ETC systems, the moments of transmission are dynamically determined by an intelligent triggering condition. This condition is reliant on factors such as system output measurements, ensuring that transmissions occur only when needed to maintain certain performance characteristics. 
The event-triggered condition should be regularly assessed and applied at every specified communication interval.

Within the event-triggered condition framework, it functions as a "fully distributed" system, operating autonomously without reliance on any central communication structure. In such systems, when each agent autonomously determines the timing of its state information broadcast, there is a dual benefit: both the control effort and the network load are significantly decreased.

\vspace{0.1in}
\subsubsection{\textbf{Control-aware Triggering}}
Adopting a strategy where decisions are made in response to the present states of the control system leads to what is known as a control-aware approach. In this approach, control systems operate moderately when in favorable states but switch to more active modes, prioritizing data transmission, in less favorable conditions. The thresholds for transmission are calibrated to ensure that each vehicle can achieve stability objectives and make transmission choices aimed at meeting performance criteria, all while lowering the overall rate of transmission. Consequently, the value of $\tau_k$ in \ref{ETC_form} for control-aware triggering is shaped as follows:
\begin{equation}
\tau_k=\inf _{t>t_k}\left\{t - t_k \mid \|\mathcal{C}_i\| \geq \beta \right\}, \quad \text { for } t \geq 0 .
\end{equation}
where $\mathcal{C}_i$ is the cost function in \ref{multi_cost}.

\section{Experimental Results}
\noindent In our experimental setup, we treated the Packet Error Rate (PER) as an independent and identically distributed $i.i.d.$ random variable with values of $0$ (representing ideal communication) and $0.6$ (indicating a 60\% packet loss scenario) to evaluate the impact of communication loss on CACC performance. The simulations were conducted with a step size of $100ms$, matching with the periodicity of communication. To assess the effectiveness and practical applicability of the proposed strategy, we conducted simulations on a platoon comprising $N_v = 10$ vehicles. For the implementation of the optimization problem, we utilized the CVXPY package in Python \cite{cvxpy}, and the Gurobi optimization package served as the solver \cite{gurobi}.

To implement the event-triggering condition, each vehicle is required to keep a record of the state at its latest event-triggered moment and to constantly observe its current state. In our approach, we employed an All-Predecessor-Leader-Following (APLF) topology. Our strategy involves establishing a link between the communication patterns and the performance of the platoon. This relationship depends heavily on the amount of vehicle information transmitted accurately and the number of nodes that successfully receive this information. We analyzed the platoon's behavior as the threshold values for the proposed triggering condition were adjusted within a specific range, focusing on the aspects of vehicle efficiency and safety.

\subsection{\textbf{Implementation Aspects}}
\noindent The specifics of the parameters employed in our simulations are detailed in Table \ref{table1}. In each scenario, which lasts for $60s$, the platoon's goal is to keep a constant gap time of $0.6s$ with the vehicle ahead. During each instance of transmission, every cooperative vehicle utilizes its five latest velocity readings, recorded at regular $100ms$ intervals, to develop a GP model and derive the parameter set $\Theta_{n}=\{\gamma_{n},\gamma_{n, noise}\}$. After establishing the GP parameters, the transmitting vehicle communicates these model parameters, its five latest velocity measurements, current position, and acceleration, all accompanied by their respective time stamps.

Moreover, the transmission packet also carries the vehicle's MPC forecast of the next $10$ velocity values (as per parameter $N$ in Table \ref{table1}). Every $100ms$, cooperative vehicles refresh their data about the preceding vehicles, using either recent direct communications or GP model predictions. This information feeds into the MPC, adjusting control decisions. The control module calculates the ego vehicle's optimal predicted states and, upon a triggering event, sends the current state and predicted velocity trajectory to the networking module for broadcasting.

\begin{figure}[t]
  \centering
    \includegraphics[width=1\linewidth]{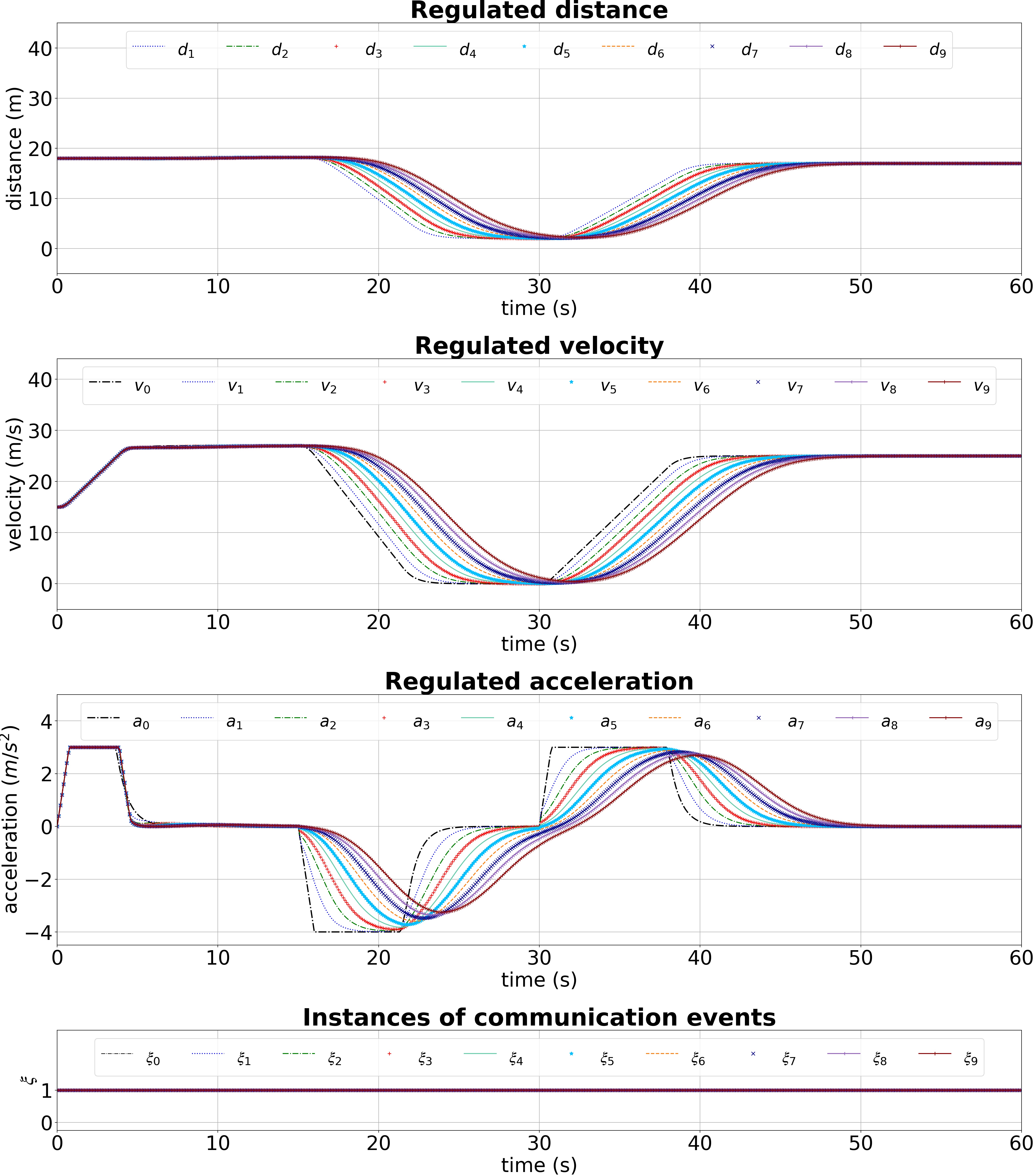}
  \caption{Functioning of CACC system using TTC with a PER of 0, and a constant communication rate at 10 Hz.}
  \label{fig:TTC_0}
\end{figure}

\begin{figure}[t]
  \centering
    \includegraphics[width=1\linewidth]{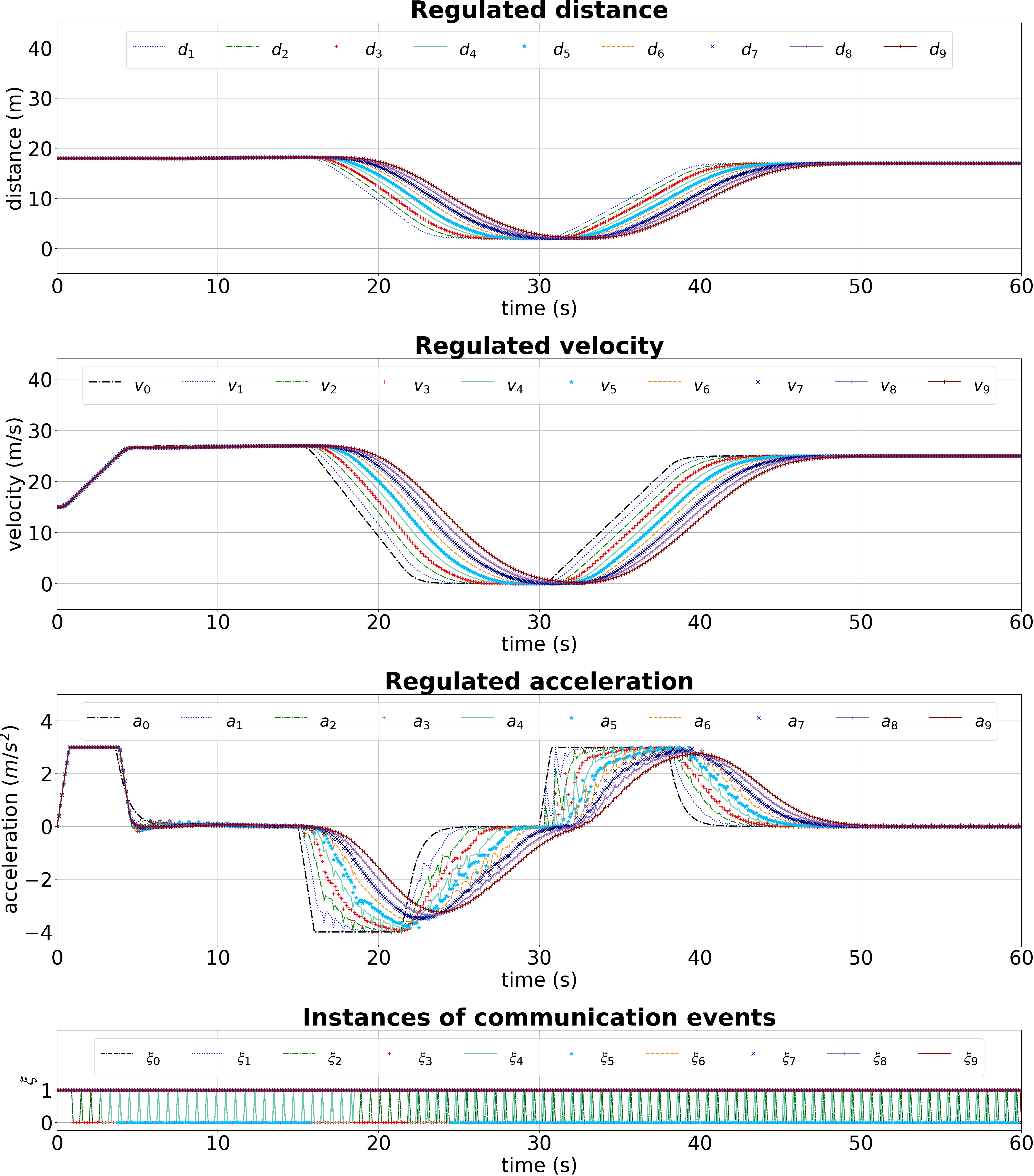}
  \caption{Functioning of the CACC system using control-aware triggered ETC, with a PER of 0, a threshold at level 6, and an average communication frequency set at 5.28 Hz.}
  \label{fig:control_ideal}
\end{figure}

\begin{table}[b]
\caption{Parameters for modeling and optimization employed in the simulations.}
\centering
\renewcommand{\arraystretch}{1.1}
\begin{tabular}{|p{1.3cm}|p{1.2cm}||p{1.3cm}|p{1.2cm}|} 
 \hline
 Parameter & Value & Parameter & Value\\ [0.5ex] 
 \hline\
 $N$ &  $10$ & $t_s$ & $0.1\,s$
 \\\
 $l^v_n$ & $5\,m$ & $d^s_n$ & $2\,m$ \\\
 $a_n^{max}$ & $ 3\,m/s^2$  & $a_n^{min}$ & $ -4\,m/s^2$\\\
 $u_n^{max}$ & $ 3\,m/s^2$  & $u_n^{min}$ & $ -4\,m/s^2$\\ 
 $f_n$ & $10\,s^{-1}$ &  & \\ 
 \hline
\end{tabular}
\label{table1}
\end{table}

\subsection{\textbf{Analysis and Results}}
\noindent In our study, we utilize the outcomes from a TTC scheme, which sets transmission times at a constant rate of $10Hz$, as a baseline to evaluate the efficacy and communication resource management of our proposed ETC approach. The average transmission rate, indicative of network resource consumption, is compared to assess the effectiveness of control-aware triggering. The distance error is calculated as the absolute difference, in meters, between the actual and ideal inter-vehicle spacing. Moreover, the differences in maximum and minimum speeds and accelerations among all platoon members at each time step are valuable metrics for assessing traffic flow dynamics and the effectiveness of the CACC system. These disparities are referred to as speed difference and acceleration difference. In an ideal $10Hz$ TTC scenario, the smallest errors recorded for mean absolute spacing error, speed difference, and acceleration difference are $\textbf{0.302}m$, $\textbf{4.693}m/s$, and $\textbf{1.257}m/s^{2}$ respectively (refer to Figure \ref{fig:TTC_0}). These figures represent the minimum errors achieved through our methodology.

\begin{table}[b]
\caption{Statistics for Control-aware Triggering with varying threshold levels}
\begin{center}
\renewcommand{\arraystretch}{1}
\begin{tabular}{|m{1.5cm}|m{1.37cm}|m{0.94cm}|m{0.94cm}|m{0.94cm}|} 
 \hline
 Thresholding level & Mean communication rate [$Hz$] & Mean Spacing error [$m$] & Mean Speed difference [$m/s$] & Mean Acceleration difference [$m/s^{2}$]\\ [0.6ex]
\hline $200$ & $7.52$ & $0.304$ & $4.697$ & $1.252$ \\
\hline $300$ & $6.82$ & $0.306$ & $4.698$ & $1.252$ \\
\hline $400$ & $6.72$ & $0.306$ & $4.698$ & $1.253$ \\
\hline $500$ & $6.04$ & $0.308$ & $4.699$ & $1.251$ \\
\hline $600$ & $5.99$ & $0.308$ & $4.700$ & $1.251$ \\
\hline $700$ & $5.28$ & $0.309$ & $4.700$ & $1.250$ \\
\hline
\end{tabular}
\label{tab:results}
\end{center}
\end{table}

\begin{figure}[t]
  \centering
    \includegraphics[width=1\linewidth]{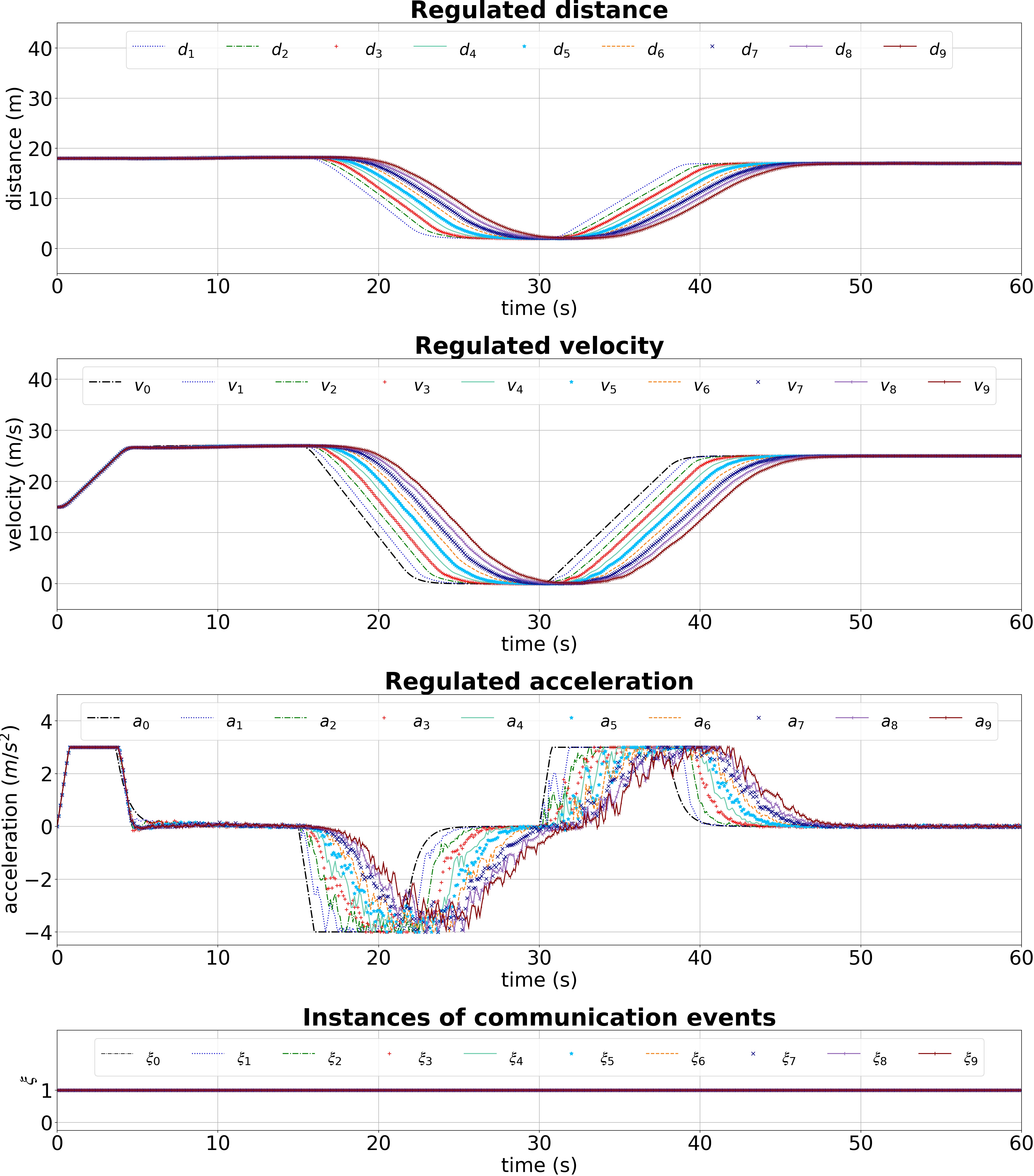}
  \caption{Functioning of the CACC system with TTC, with a PER of 0.6, and a constant communication rate at 10 Hz.}
  \label{fig:TTC_0.6}
\end{figure}

\begin{figure}[t]
  \centering
    \includegraphics[width=1\linewidth]{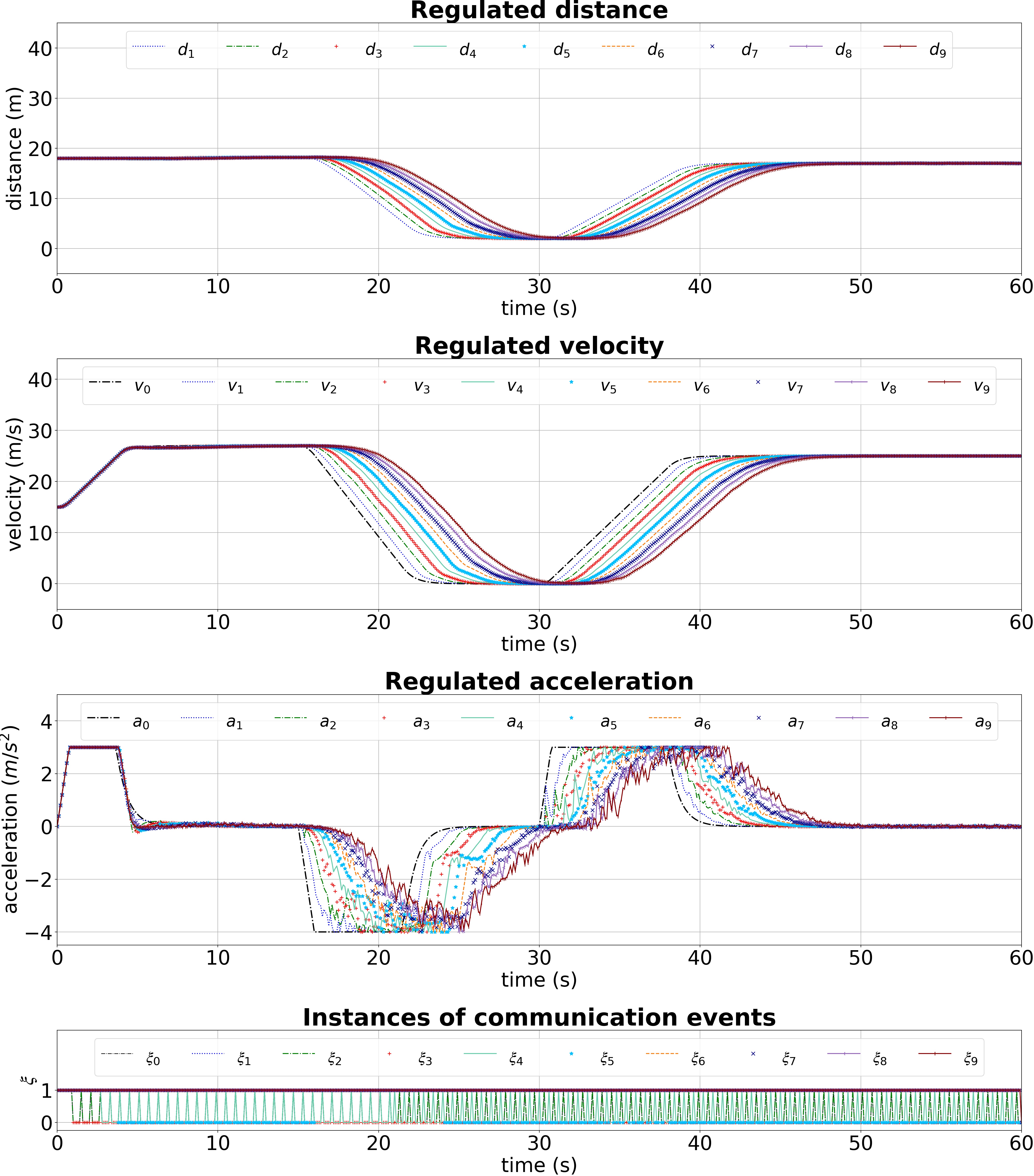}
  \caption{Functioning of the CACC system using control-aware triggered ETC, with a PER of 0.6, a threshold at level 6, and an average communication frequency set at 5.28 Hz.}
  \label{fig:control_aware_0.6}
\end{figure}

Figures \ref{fig:TTC_0}, \ref{fig:control_ideal}, \ref{fig:TTC_0.6}, and \ref{fig:control_aware_0.6} each consist of four subplots. The first row of subplots displays the distance of each vehicle from the one ahead ($d_{n}(t)$), the second row shows each vehicle's velocity ($v_{n}(t)$), the third row presents acceleration data for each vehicle, and the final row marks each moment ($\xi_{n}(t)$) a vehicle transmits information to the vehicles following it. It's important to note that for the leader of the platoon, $d_{0}(t)$ is not defined as it has no preceding vehicle, so this data is omitted from the first subplots of these figures. Despite a marked decrease in communication frequency with the ETC scheme, its response patterns are similar to those observed with the TTC scheme. This illustrates that communication frequency can be substantially reduced without compromising control performance.

Figure \ref{fig:control_ideal} illustrates the optimal scenario simulation with a level 6 threshold. Here, we only consider the impact of the ETC scheme under ideal communication conditions ($PER=0$). In contrast, Figure \ref{fig:control_aware_0.6} represents the simulation's most challenging scenario, combining the effects of both the ETC scheme and the highest Packet Error Rate (PER), under the most strictest threshold settings. Unlike the smooth acceleration profile in Figures \ref{fig:TTC_0} and \ref{fig:control_ideal}, acceleration in other figures shows variability due to the occasional lack of precise information, either from packet loss or untriggered transmissions. Nevertheless, vehicles following the proposed communication strategy can maintain safe following distances. The control-aware triggering thresholds are set in six evenly spaced stages ranging from $200$ to $700$, as detailed in Table \ref{tab:results}.

Figure \ref{fig:control_ideal} depicts a scenario where communication is initiated based on the present state of the control system. Notably, the vehicles at the end of the platoon tend to experience more challenging control conditions, as they need to adjust for the errors of the vehicles ahead to maintain string stability in the platoon. Consequently, these following vehicles, particularly the later members, engage in more frequent transmissions. For example, the communication events for $\xi_{8}(t)$ and $\xi_{9}(t)$ are consistently active.

Table \ref{tab:results} illustrates the balance that can be achieved between control performance and data transmission rate by adjusting the trigger level. To provide a more accurate representation of performance, each value in the table is the average of 70 simulation runs. The table presents the average values for spacing error ($m$), speed difference ($m/s$), and acceleration difference ($m/s^{2}$). The implementation of the ETC policy creates a compromise between the effectiveness of control and the frequency of communication. It is evident that reducing the trigger level results in lower errors but necessitates a higher rate of data transmission. Extended durations between events lead to notable errors in performance. Consequently, increasing the threshold level tends to adversely affect control performance.

\section{Conclusion}
\noindent The scalability of distributed time-triggered communication methods diminishes as the number of devices on a shared network surpasses previous capacities. This limitation calls for a move from regular, periodic communication methods to more adaptive, opportunistic approaches, such as the one proposed in this paper. Overutilization of communication resources can adversely affect their dependability. Therefore, this article introduces a resource-efficient CACC communication strategy, aimed at reducing the use of communication resources compared to traditional TTC techniques, while still preserving the system's performance. Alternatively, this approach can enhance system performance at a constant communication rate. Additionally, to prevent Zeno behavior, the design ensures that the minimum times between events always have a positive lower limit.

Furthermore, our approach integrates MBC with ETC to develop a communication strategy for distributed multi-agent coordination. In this system, each agent's decision to transmit new measurements across the network is primarily based on the deviation between its current state or model and the state or model at the time of its last transmission. It is necessary only to periodically verify and implement the event-triggered condition at each designated time for communication. The simulation results demonstrate the feasibility of an ETC system that not only effectively reduces network load by $47\%$ compared to TTC but also minimally impacts control performance, such as less than $1\%$ in speed deviation.
\vspace{0.1in}
\balance
\bibliography{main.bib}{}
\bibliographystyle{unsrt}
\end{document}